# PINO-MBD: Physics-informed Neural Operator for Solving Coupled ODEs in Multi-body Dynamics

Wenhao Ding[1], Qing He[1†], Hanghang Tong[2], Ping Wang[1]
[1]MOE Key Laboratory of High-speed Railway Engineering, Southwest Jiaotong University
[2]University of Illinois at Urbana-Champaign
whding@my.swjtu.edu.cn, qhe@swjtu.edu.cn, wping@home.swjtu.edu.cn
htong@illinois.edu

## ABSTRACT

In multi-body dynamics, the motion of a complicated physical object is described as a coupled ordinary differential equation system with multiple unknown solutions. Engineers need to constantly adjust the object to meet requirements at the design stage, where a highly efficient solver is needed. The rise of machine learning-based partial differential equation solvers can meet this need. These solvers can be classified into two categories: approximating the solution function (Physics-informed neural network) and learning the solution operator (Neural operator). The recently proposed physics-informed neural operator (PINO) gains advantages from both categories by embedding physics equations into the loss function of a neural operator. Following this state-of-art concept, we propose the physics-informed neural operator for coupled ODEs in multi-body dynamics (PINO-MBD), which learns the mapping between parameter spaces and solution spaces. Once PINO-MBD is trained, only one forward pass of the network is required to obtain the solutions for a new instance with different parameters. To handle the difficulty that coupled ODEs contain multiple solutions (instead of only one in normal PDE problems), two new physics embedding methods are also proposed. The experimental results on classic vehicle-track coupled dynamics problem show state-of-art performance not only on solutions but also the first and second derivatives of solutions.

## CCS CONCEPTS

•Applied computing ➙ Physical sciences and engineering ➙ Engineering ➙ Computer-aided design

## 1 Introduction

### 1.1 Machine learning-based PDE solvers

Almost all physics and engineering problems can be described with partial differential equations (PDE). Using machine learning-based methods to solve PDEs has become a popular research topic in recent years. Compared with classic solvers, ML-based methods have shown the lower computational cost and better accuracy in certain cases [1-3]. This has given rise to applied research in many fields such as physics [4-5], chemistry [6], and engineering [7-8].

At present, the mainstream theories can be mainly divided into two categories. The first one utilizes deep neural networks (DNN) as a solver to learn the solution function of a specific PDE. One of the most representative works is the physics-informed neural network (PINN) [2]. PINN uses automatic differential operation to obtain the partial differential term of the solution. Then the PDE losses can be computed and embedded into the loss function of the DNN. However, the theory itself still faces some inevitable problems. Firstly, PINN is a substitute for traditional numerical solvers. The training process of the DNN is the solution process of the PDE, and the training process is often sensitive to the selection of hyperparameters. When the PDE parameters change, one needs to retrain and fine-tune the PINN. Secondly, the value of PINN only exists in cases where its computational cost is lower than the mathematic solver. Therefore, PINN is not practical when dealing with a large number of PDE problems with lower complexity.

To tackle the main defect of PINN, the theory of neural operators was proposed for parametric PDEs [9]. This recently proposed theory aims to learn the solution operator for a family of PDEs, which is the mapping between the input parameter space and the output solution space. Once the neuron operator is trained, only one forward pass of the network is required to obtain the solution for a new instance of the parameter. Recently a series of neuron operators have been proposed, including Graph Neural Operator (GNO) [9], Low-rank Neural Operator (LNO) [10], Multipole Graph Neural Operator (MGNO) [11], and Fourier Neural Operator (FNO) [12], all of which are derived from graph neural networks (GNN) [13-14]. FNO utilizes the calculus properties of the fast Fourier transform to perform convolution operation of GNN, which is the state-of-art model among them. On the other hand, the DeepONet proposed at the same time shares the same concept with the neural operator [15]. The difference is reflected in the network architecture. The DeepONet processes the data on two parallel networks (branch net and trunk net) to handle input and output functions separately. Then the mapping operator for parametric PDEs can be learned by combing the output of the two networks. This model has derived a few applied research directions as well, including the inference of the electroconvection Multiphysics fields [16] and prediction for finite-rate chemistry [17]. Recently,

Kovachki has proven that the standard DeepONet can be mathematically regarded as a special case of neural operators [10], which makes the theory of neuron operators more general.

Both PINN and neuron operator have their own advantages and disadvantages. The former is a one-shot solver, and the learning ability of the DNN is enhanced by embedding the physics information. The latter deals with parametric PDE problems but neglects the physics information. Obviously, a very simple and natural idea is to embed the physical equation in the loss function of the neuron operator just like PINN, so as to get the advantages of both. Driven by this idea, Li proposed Physics-informed Neural Operator (PINO) [18], and Wang proposed Physics-informed DeepONet [19]. Both studies found that neuron operators can achieve better performance on smaller datasets by embedding physics system equations. However, when applied in multi-body dynamics, PINO suffers from many problems, which will be described in Section 1.2.

## 1.2 PINO for coupled ordinary differential equations in multi-body dynamics

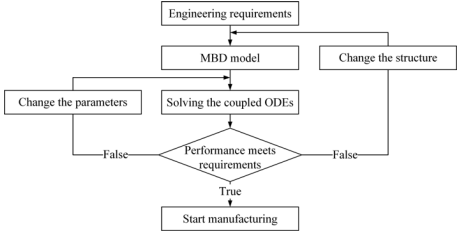

**Figure 1: Flow chart for typical MBD engineering problem**

Multi-body dynamics (MBD) models a complex physical object by transforming it into a system composed of multiple interconnected bodies [20-21]. Each degree of freedom (DOF) for the bodies is described with an ordinary differential equation (ODE). And each ODE in the system is coupled. All the ODEs are combined to form the coupled ordinary differential equation system (CODES) for the entire system. MBD has been applied in many practical engineering problems, such as aerospace vehicles, locomotives, automobiles, robotics, etc. As shown in Figure 1, for a typical MBD system design problem, engineers need to continuously change the structure and parameters of the system according to actual needs until the design meets the requirements [20]. Obviously, there is a huge demand for solving parametric CODESs in this process. However, the study on this matter in the engineering area is currently still at the stage of using ordinary DNN [25-26] and time series forecasting models such as LSTM [27]. On the other hand, all existing data-driven methods for CODES are based on PINN and therefore cannot handle parametric CODES [6, 28, 29]. Following the concept of PINO, we aim to propose a physics-informed neural operator PINO-MBD for CODES in MBD. Our contributions mainly lie in two perspectives.

(1). Common physical PDEs usually contain only one unknown solution, such as the classic Burgers Equation and Navier-Stokes Equation. A single CODES contains multiple unknown solutions and parameters, and the solutions are coupled with each other. We propose two new methods to embed the CODES into the loss function. We will show in section 4 that the existing loss function embedding method used by PINO and PINN cannot handle the problem of parametric CODESs.

(2). We propose a PINO for the MBD problem for the first time. We conduct experiments on a widely used MBD problem, vehicle-track coupled dynamics. Results show that PINO-MBD can not only ensure the accuracy of the solution, but also the accuracy of the derivatives of the solution.

## 2 Preliminaries and MBD problem settings

### 2.1 Terminology and MBD Fundamentals

Almost all multibody dynamics models can be assembled with three types of components: bodies, force elements, and excitations. As shown in Figure 2, the bodies can be rigid or flexible. When subjected to forces, the shape of the rigid body and the relative positions of its internal points remain unchanged. It is usually used to model objects with negligible deformation during motion. Conversely, a flexible body can deform during motion, and the relative positions of internal points can change. It is usually used to describe objects whose deformation cannot be ignored during motion.

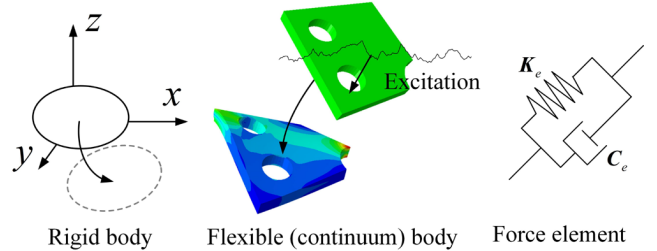

**Figure 2: General MBD Composition**

Force elements are defined as the interconnections between bodies. There are various forms of force elements. Some force elements limit the relative motion of connected objects, while most do not, such as the classic linear spring-damper model (KV model). A system composed of force elements and multiple bodies is called a multibody system. Excitation is defined as the action of objects outside the multibody system on bodies in the system. In practical engineering, multi-body systems can well simulate the dynamic behavior of complex mechanical systems such as robots, manipulators, and high-speed trains, while external excitation is inevitable. For example, the robot will be disturbed by the external force of the environment, and the mechanical arm needs to move various heavy objects. In another MBD case, the train movement is subjected to rail irregularities constantly during operation.

### 2.2 MBD Problem settings

In order to learn the relationship between input functions and the multiple response functions of a multi-body system with neural operators, we need to construct the system equations that can be

# PINO-MBD: Physics-informed Neural Operator for Solving Coupled ODEs in Multi-body Dynamics

embedded in a deep network. Without loss of generality, we take a general multi-body system in Figure 3 as an example to illustrate, where rigid/flexible bodies, force element, and external excitations are all considered.

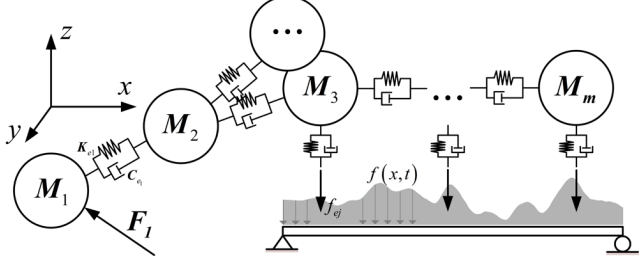

**Figure 3: General Topology for a Multi-body System**

**Table 1: Main Symbols**

| Symbol | Definition and Description |
|---|---|
| $M_i$ | the mass matrix for the i-th rigid body |
| $X_i$ | the displacement vector for the i-th rigid body |
| $F_i$ | the external force vector acting on the i-th rigid body |
| $K_{ej}$ | the stiffness matrix for force element j |
| $C_{ej}$ | the damping matrix for force element j |
| $E, I, \rho, A, l$ | elastic module, moment of inertia, density, section cross-sectional area, and the length of the beam |
| $w(x,t)$ | the deflection function of the flexible beam |
| $W_n(x)$ | the nth mode shape function (function solution in space domain) |
| $T(x)$ | function solution in time domain |
| $\omega_n$ | the nth natural frequency of the flexible beam |
| $f(x,t)$ | The force function acting on the flexible beam |
| $q_n(t)$ | the generalized coordinate of the nth mode function |
| $n_r$ | total number of degree of freedoms of rigid bodies |
| $n_f$ | total number of mode shape functions considered for the flexible beam |
| $x, y, z, t$ | Space coordinates and time |

For the classic KV model, the output of the force element is linear to the relative displacement and first-order derivative (velocity) of the bodies at both ends. Taking the No. 1 force element in Figure 2 as an example, the force element output $f_{e1}$ can be written as:

$$f_{e1} = K_{e1}(X_1 - X_2) + C_{e1}\left(\frac{d}{dt}X_1 - \frac{d}{dt}X_2\right) \quad (1)$$

As mentioned earlier, bodies are classified into rigid bodies and flexible bodies. Each degree of freedom of a rigid body is described by an ordinary differential equation (ODE), which can be obtained using Newton's second law, D'Alembert's principle, or the energy method. Taking the first rigid body in Figure 3 as an example, its motion equations in all directions can be described by matrices as:

$$M_1 \frac{d^2}{dt^2} X_1 + C_e\left(\frac{d}{dt}X_1 - \frac{d}{dt}X_2\right) + K_e(X_1 - X_2) = 0 \quad (2)$$

Flexible bodies are usually described with continuum mechanical models. Physical equations for different mechanical models are different. However, they are all high-order PDEs with respect to time and space. In engineering, the common processing method is to first obtain the mode shape function of the continuum through the free vibration PDE, and then use the mode superposition method to convert the forced vibration PDE into a set of low-order ODEs. From there they can be coupled with other ODEs to form the complete CODES for the entire multi-body system. We will take the simply supported beam in Figure 3 as an example to clarify, the free vibration equation of the continuum is a 4th-order PDE:

$$EI\frac{\partial^4 w(x,t)}{\partial x^4} + \rho A \frac{\partial^2 w(x,t)}{\partial t^2} = 0 \quad (3)$$

The free-vibration solution can be found using the method of separation of variables as

$$w(x,t) = W(x)T(t) \quad (4)$$

Substituting Eq. (4) into the PDE (3) leads to

$$\frac{c^2}{W(x)}\frac{d^4 W(x)}{dx^4} = -\frac{1}{T(t)}\frac{d^2 T(t)}{dt^2} = \omega^2 \quad (5) \qquad c = \sqrt{\frac{EI}{\rho A}} \quad (6)$$

$$\frac{d^4 W(x)}{dx^4} - \beta^4 W(x) = 0 \quad (7) \qquad \frac{d^2 T(t)}{dt^2} + \omega^2 T(t) = 0 \quad (8)$$

$$\beta^4 = \frac{\omega^2}{c^2} = \frac{\rho A \omega^2}{EI} \quad (9)$$

where $\omega$ is referred to as the natural frequency and $W(x)$ the mode shape function. There are an infinite number of mode functions that satisfy (5), each corresponding to a different $\omega$. And different mode shape functions satisfy the orthogonality condition:

$$\int W_i W_j dx = 0, \quad x \in [0,l] \quad (10)$$

When the beam is subjected to external excitation, the forced vibration PDE is formed as

$$EI\frac{\partial^4 w(x,t)}{\partial x^4} + \rho A \frac{\partial^2 w(x,t)}{\partial t^2} = f(x,t) \quad (11)$$

The solution can be determined using the mode superposition principle. For this, the deflection of the beam is assumed as

$$w(x,t) = \sum_{n=1}^{\infty} W_n(x) q_n(t) \quad (12)$$

where $W_n$ is the $n$th normal mode function satisfying (7), and $q_n(t)$ is referred to as the generalized coordinate in the $n$th mode. By substituting Eq. (12) into (11), we obtain

$$\sum_{n=1}^{\infty} \omega_n^2 W_n(x) q_n(t) + \sum_{n=1}^{\infty} W_n(x)\frac{d^2 q_n(t)}{dt^2} = \frac{1}{\rho A} f(x,t) \quad (13)$$

By multiplying Eq.(13) with $W_m(x)$, integrating over $[0,l]$, and using Eq. (4), we have

$$\frac{d^2 q_n(t)}{dt^2} + \omega_n^2 q_n(t) = \frac{1}{\rho A b} Q_n(t) \quad (14) \qquad Q_n(t) = \int_0^l f(x,t) W_n(x) dx \quad (15)$$

where $Q_n(t)$ is referred to the generalized force acting on $q_n(t)$

xThe constant $b$ is obtained with Eq. (16).

$$b = \int_0^l W_n^2(x)dx \qquad (16)$$

So far, the forced vibration Eq. (11) has been transformed from a 4th-order PDE to $n_f$ second-order ODEs, each of which describes the motion of one mode of the continuum. By simply assembling them with those that describe the motions of rigid bodies, we have the CODES of the entire multi-body system:

$$M_\Theta \frac{d^2}{dt^2}X_\Theta + C_\Theta \frac{d}{dt}X_\Theta + K_\Theta X_\Theta = F_\Theta$$

$$X_\Theta = \begin{bmatrix} X_1^T \cdots X_{\mu_r}^T q_1 \cdots q_{\mu_f} \end{bmatrix}^T \qquad (17)$$

$$F_\Theta = \begin{bmatrix} F_1^T \cdots F_{\mu_r}^T Q_1 \cdots Q_{\mu_f} \end{bmatrix}^T$$

where $M_\Theta$, $C_\Theta$ and $K_\Theta$ are the mass, damping and stiffness matrix of the multi-body system, respectively. More details of the formulas above can be found in [22].

A few points need to be emphasized,

(1). There is an infinite number of modes for the continuum, each corresponding to one natural frequency. However, in practical applications, we only need to consider a certain number of modes ($n_f$) to ensure that the maximum natural frequency is high enough.

(2). For other continuum models, such as strings, rods, and membranes, although the physical equations are different, the solution methods are similar. Research in the field of mechanics has provided a large number of methods that can solve Eq. (5) to obtain the mode shape function. In this case, we mainly use the Ritz method [22] with high precision. In practical applications, the mode shape functions for many complex flexible body structures cannot be found through physical derivation. In this case, one can conduct finite element modal analysis to acquire $W_n(x)$.

(3). The generalized coordinate response is solved in the CODES. One can simply apply Eq. (10) to reconstruct the solution of the continuum.

## 2.3 Learning the solution operator

As described in Section 2.2, any multi-body dynamical system can be described with CODES (17):

$$M_\Theta \frac{d^2}{dt^2}X_\Theta + C_\Theta \frac{d}{dt}X_\Theta + K_\Theta X_\Theta = F_\Theta \qquad (17)$$

where $M_\Theta$, $C_\Theta$ and $K_\Theta$ can be constructed with the system's physical parameters. Therefore CODES (17) can be further described as

$$\Phi_i(X_\Theta, a, F_\Theta(i,:)) = 0 \quad i=1,2,\cdots \mu_r+\mu_f, \quad in(0,T)$$
$$X_\Theta(0) = G_\Theta \qquad (18)$$

where $a(j,:) \in A_j$ $j=1,2,\cdots n_p$ is the physical parameter vector for the CODES, and $n_p$ is the number of parameters. $X_\Theta(i,:) \in U_i$ is the unknown solution of the $i$th ODE in the CODES. $G_\Theta$ is the initial condition of the CODES. $\Phi_i$ is a possibly non-linear ordinary differential operator for the $i$th ODE in the CODES with three sets of Banach spaces

$$\{(U_1, U_2, \cdots, U_{\mu_r+\mu_f}), (A_1, A_2, \cdots, A_{\mu_p}), F_i\}.$$

This formulation gives rise to the solution operator

$$\mathcal{G}^\dagger : \{(A_1, A_2, \cdots, A_{\mu_p}), (F_1, F_2, \cdots, F_{\mu_r+\mu_f})\} \to (U_1, U_2, \cdots, U_{\mu_r+\mu_f}) \qquad (19)$$

To emphasize, each $\Phi_i$ is related to all $U_i$ and $A_i$ due to the coupled nature of all ODEs. This leads to two difficulties in approximating operator $\mathcal{G}^\dagger$.

(1). Unlike existing researches such as PINO and PINN, operator $\mathcal{G}^\dagger$ maps to multiple solution spaces ($n_r + n_f$ in total) instead of one. In real MBD applications, the number usually reaches dozens or even hundreds.

(2). Different from existing research that only focuses on the solution of the physical equation, in practical MBD engineering problems, more attention is paid to the derivatives of the unknown solutions, that is $\frac{d^2}{dt^2}X_\Theta$ and $\frac{d}{dt}X_\Theta$. For example, sound in nature is usually associated with the first derivatives of the solutions (velocity), and the human body is more sensitive to the second derivatives of the solutions (acceleration).

Given a CODES defined in (17) and the corresponding solution operator $\mathcal{G}^\dagger$, one can use a neural operator $\mathcal{G}_\theta$ with parameters $\theta$ as a surrogate model to approximate $\mathcal{G}^\dagger$. We assume a dataset $\{a_k, F_{\Theta k}, X_{\Theta k}\}_{k=1}^N$ is available, where $\mathcal{G}^\dagger(a_k, F_{\Theta k}) = X_{\Theta k}$ and each component of $a_k$ and $F_{\Theta k}$ are i.i.d sampled from some distribution $\mu_a$ and $\mu_F$ supported on corresponding spaces. In this case, one can optimize the solution operator by minimizing the empirical data loss on a given data pair

$$L_{data}(X_{\Theta k}, \mathcal{G}^\dagger(a_k, F_{\Theta k})) = \|X_{\Theta k} - \mathcal{G}^\dagger(a_k, F_{\Theta k})\|^2 = \int_0^T |X_{\Theta k}(t) - \mathcal{G}^\dagger(a_k, F_{\Theta k})(t)|dt \qquad (20)$$

The operator data loss is defined as the average error across all possible inputs

$$J_{data}(\mathcal{G}_\theta) = \|\mathcal{G}^\dagger - \mathcal{G}_\theta\|^2 = \mathbb{E}[L_{data}(a_k, F_{\Theta k}, \theta)] \approx \frac{1}{N}\sum_{k=1}^N \int_0^T |X_{\Theta k}(t) - \mathcal{G}^\dagger(a_k, F_{\Theta k})(t)|dt \qquad (21)$$

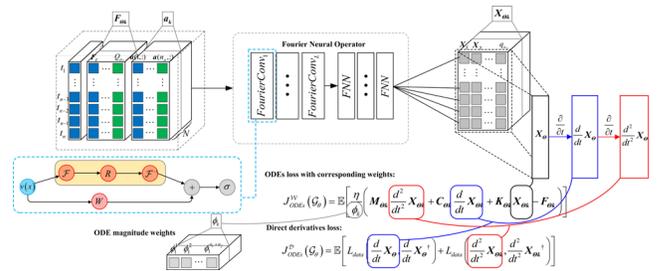

**Figure 4: Network structure of PINO-MBD**

## 3 Physics-informed neural operator for MBD

### 3.1 Fourier neural operators

The neural operators proposed in [9, 10], is formulated as a generalization of standard deep neural networks to operator setting. Following the basic mathematical principles of Graph Neural Networks (GNN), Neural operator composes a linear integral operator $\mathcal{K}$ with pointwise non-linear activation function $\sigma$ to approximate highly non-linear operators.

**Definition 1 (Neural operator $\mathcal{G}_\theta$)** Define the neural operator

$$\mathcal{G}_\theta := \mathcal{Q} \circ (W_L + \mathcal{K}_L) \circ \cdots \circ \sigma(W_1 + \mathcal{K}_1) \circ \mathcal{P} \quad (22)$$

where $\mathcal{P}: \mathbb{R}^{d_a} \to \mathbb{R}^{d_1}$, $\mathcal{Q}: \mathbb{R}^{d_L} \to \mathbb{R}^{d_u}$ are the pointwise neural networks that encode the lower dimension function into higher dimensional space and decode the higher dimension function back to the lower-dimensional space. The model stack $L$ layers of $\sigma(W_l + \mathcal{K}_l)$ where $W \in \mathbb{R}^{d_{l+1} \times d_l}$ are pointwise linear operators (matrices), $\mathcal{K}_l : \{D \to \mathbb{R}^{d_l}\} \to \{D \to \mathbb{R}^{d_{l+1}}\}$ are integral kernel operators, and $\sigma$ are fixed activation functions. The parameters $\theta$ consist of all the parameters in $\mathcal{P}$, $\mathcal{Q}$, $W_l$, $\mathcal{K}_l$.

**Definition 2 (Fourier convolution operator $\mathcal{K}$)** Define the Fourier convolution operator

$$(\mathcal{K}v_t) = \mathcal{F}^{-1}(R \cdot (\mathcal{F}v_t)) \quad (23)$$

where $R$ is part of the parameter $\theta$ to be learned. $\mathcal{F}$ and $\mathcal{F}^{-1}$ denote the forward and inverse Fourier transform, respectively.

### 3.2 Physics embedding

As described in Section 2.2, in order for neural operator to learn the physics nature, we need to embed the CODES (17) into the loss function of the deep neural network. Similar to (21), the operator ODEs loss can be defined as:

$$J_{ODEs}(\mathcal{G}_\theta) = \mathbb{E}\left[L_{ODEs}(a_k, F_{\Theta k}, \theta)\right]$$
$$= \mathbb{E}\left[M_{\Theta k}\frac{d^2}{dt^2}X_{\Theta k} + C_{\Theta k}\frac{d}{dt}X_{\Theta k} + K_{\Theta k}X_{\Theta k} - F_{\Theta k}\right] \quad (24)$$

PINN, PINO, and Physics-informed DeepONets all employ Eq. (24) to embed the physical equations. However, Eq. (24) cannot handle CODES (17). In this section, we propose two new embedding methods $J^W_{ODEs}(\mathcal{G}_\theta)$ and $J^D_{ODEs}(\mathcal{G}_\theta)$ to deal with this challenge. The first one improves Eq. (24) by introducing ODE magnitude weight factors while the other directly provides the network with ground truth for the derivatives of the unknown solutions.

*3.2.1 ODEs losses with ODE magnitude weight factors.* Unlike most PDE problems, one CODES contains multiple coupled ODEs. Since different ODE represents different physical motions, the magnitude for each equation varies greatly. We will clarify the phenomenon on an engineering example to be dealt with in Section 4, where the CODES of this physical system mainly contain 10 ODEs. Figure 5 shows the solutions of three ODEs (1, 2, 7) among the 10 ODEs, along with their first and second derivatives. We consider adding white noise to all signals to simulate the process of neuron operators trying to approximate the real solution. The amplitude of the white noise is set as 15% of the variance of each signal itself. It can be observed that ODE loss magnitude for different ODE in the CODES vary greatly during the learning process. One can anticipate large gradient differences on different ODEs during the training process. Therefore, if we set the corresponding weight factors for different ODEs, the gradients on different ODEs are similar in magnitude during the learning process and better performance can be expected. To further explain this countermeasure, we define the ODE magnitude weight factor as follows.

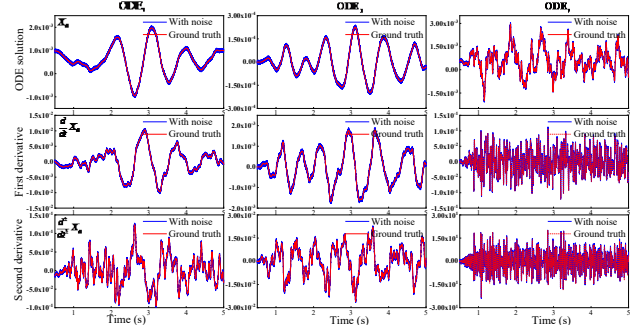

(a). Solutions & derivatives of different ODEs

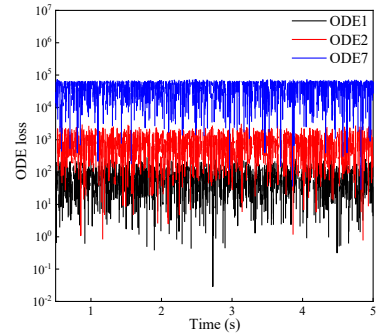

(b). Huge magnitude difference for ODE losses

**Figure 5: Huge magnitude difference for different ODE losses in one coupled ODEs**

**Definition 3 (ODE magnitude weight factor $\phi^i_k$)** Define the ODE magnitude weight factor

$$\phi^i_k = \max \Phi_i\left(X^\varepsilon_{\Theta k}, a, F_{\Theta k}(i,:)\right) \quad k = 1 \sim N, i = 1 \sim n_r + n_f$$
$$X^\varepsilon_{\Theta k}(:,i) = X_{\Theta k}(:,i) + \varepsilon\left(\mu = 0, \sigma^2 = r \cdot D(X_{\Theta k}(:,i))\right) \quad (25)$$

where $\varepsilon$ is white noise sequence, and $r$ is the sensitivity factor, meaning the acceptable signal error. In the previous paragraph, we set $r$ as 15%. In practical applications, we usually set r to be 2% for more penalties. The ODE magnitude weight factors calculated through Eq. (25) corresponds to the data pair and each ODE one-to-one, and is used together with the training data during the training process to form

$$J^W_{ODEs}(\mathcal{G}_\theta) = \mathbb{E}\left[\frac{\eta}{\phi^i_k}\left(M_{\Theta k}\frac{d^2}{dt^2}X_{\Theta k} + C_{\Theta k}\frac{d}{dt}X_{\Theta k} + K_{\Theta k}X_{\Theta k} - F_{\Theta k}\right)\right] \quad (26)$$

$r$ in Eq. (25) is the loss function target that one wishes to reduce to. We take the value of 2%.

*3.2.2 Direct derivatives losses* The solutions obtained through numerical methods naturally satisfy the CODES. For MBD problems, a series of mature numerical integration methods have been developed to solve their CODES. Whether implicit (implicit Euler method, Newmark-$\beta$) or explicit (Runge-Kutta method, Zhai-method), most numerical integration methods can acquire solutions and the derivatives of solutions at the same time. That is, we can easily obtain ground truth for derivatives of the solutions before the training process. In the existing research field, only the ground truth for solutions is provided for the neural operator while the derivatives of solutions are usually left wasted. Therefore, we hope to directly provide the ground truth of derivatives for the neural operator in the same way as the data losses.

$$J^{\mathcal{D}}_{ODEs}(\mathcal{G}_\theta) = \mathbb{E}\left[L_{data}\left(\frac{d}{dt}X_\Theta, \frac{d}{dt}X_\Theta^\dagger\right) + L_{data}\left(\frac{d^2}{dt^2}X_{\Theta k}, \frac{d^2}{dt^2}X_{\Theta k}^\dagger\right)\right] \quad (27)$$

The neural operator only outputs solutions of the CODES and their derivatives will be computed through differential operations. Li has summarized three differential operations to obtain solution derivatives: Autograd, numerical difference, and exact gradients [18]. In this paper, we mainly use numerical difference method to solve the derivatives. Numerical difference method is faster and less memory-consuming compared to Autograd. It is also more stable compared to exact gradients.

## 4 Experiment: Train-Track coupled dynamics

In this section, we will conduct experiments on the widely used vehicle-track coupled dynamics [23]. We will assemble the CODES in section 4.1. In section 4.2, we will show that PINO-MBD achieves better performance on derivatives of solutions compared with ordinary neural operators. We also show the two proposed physics embedding methods can help PINO-MBD achieve much better results than normal embedding method (24).

**Table 2: Main Symbols for the train-track coupled dynamics**

| Symbol | Definition and Description |
|---|---|
| $M_*$ | the mass of body * |
| $J_*$ | the moment of inertia of body * |
| $Z_*$ | the vertical displacement for body * |
| $\beta_*$ | the rotation angle for body * |
| $Z_r(x,t)$ | displacement of rail at position x, time t |
| $l_c, l_t$ | the characteristic length of the vehicle system |
| $q_k$ | the kth mode displacement for the flexible rail |
| NM | number of rail modes selected |
| N | number of fasteners |
| $p_j$ | wheel-rail force at jth wheelset |
| v | running speed of the train |
| $K_p, K_s$ | the stiffness of primary and secondary suspension |
| $C_p, C_s$ | the damping of primary and secondary suspension |
| $K_p, C_p$ | the stiffness and damping of rail fastener |
| $Irre_i(t)$ | rail irregularity under the ith wheelset |
| $E, I_Y, m_r, l$ | elastic module, moment of inertia, mass per mass, and the length of the rail |
| $\cdot$ | first derivative with respect to time |
| $\cdot\cdot$ | second derivative with respect to time |
| Body * | Definition and Description |
| c | car body |
| $t_j$, j=1~2 | the jth bogie |
| $w_j$, j=1~4 | the jth wheelset |

### 4.1 MBD equations settings

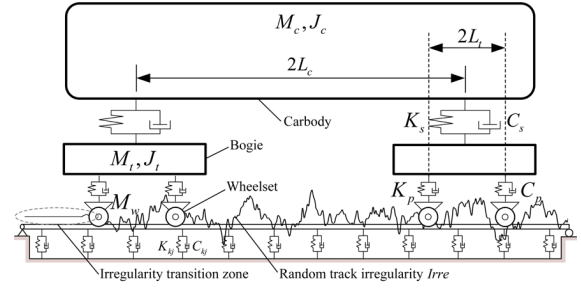

**Figure 6: Topology of the vehicle-track coupled system**

The vehicle-track coupled dynamics (VTCD) is a typical MBD problem [23]. This theory has been tested by numerous field experiments and continues to play an important role in the design and construction of high-speed railways. The schematic and topology are shown in Figure 6. The vehicle system with 10 degrees of freedom is composed of rigid bodies and force elements. And the rail is modeled with a continuous Euler beam supported on a series of force elements. Its CODES can be formed following the procedure in Section 2.2 as

$$\begin{aligned}
&M_c\ddot{Z}_c + 2C_s\dot{Z}_c + 2K_{sz}Z_c - C_{sz}(\dot{Z}_{t1} + \dot{Z}_{t2}) - K_{sz}(Z_{t1} + Z_{t2}) = M_c g \\
&J_c\ddot{\beta}_c + 2C_s l_c^2 \dot{\beta}_c + 2K_s l_c^2 \beta_c + C_s l_c(\dot{Z}_{t1} - \dot{Z}_{t2}) + K_s l_c(Z_{t1} - Z_{t2}) = 0 \\
&M_t\ddot{Z}_{t1} + (2C_p + C_s)\dot{Z}_{t1} + (2K_p + K_s)Z_{t1} - C_s\dot{Z}_c - K_s Z_c \\
&\quad -C_p(\dot{Z}_{w1} + \dot{Z}_{w2}) - K_p(Z_{w1} + Z_{w2}) + C_s l_c \dot{\beta}_c + K_s l_c \beta_c = M_t g \\
&J_t\ddot{\beta}_{t1} + 2C_p l_t^2 \dot{\beta}_{t1} + 2K_p l_t^2 \beta_{t1} + C_p l_t(\dot{Z}_{w1} - \dot{Z}_{w2}) + K_p l_t(Z_{w1} - Z_{w2}) = 0 \\
&M_t\ddot{Z}_{t2} + (2C_p + C_s)\dot{Z}_{t2} + (2K_p + K_s)Z_{t2} - C_s\dot{Z}_c - K_s Z_c \\
&\quad -C_p(\dot{Z}_{w3} + \dot{Z}_{w4}) - K_p(Z_{w3} + Z_{w4}) + C_s l_c \dot{\beta}_c + K_s l_c \beta_c = M_t g \\
&J_t\ddot{\beta}_{t2} + 2C_p l_t^2 \dot{\beta}_{t2} + 2K_p l_t^2 \beta_{t2} + C_p l_t(\dot{Z}_{w3} - \dot{Z}_{w4}) + K_p l_t(Z_{w3} - Z_{w4}) = 0 \\
&M_w\ddot{Z}_{w1} + C_p(\dot{Z}_{w1} - \dot{Z}_{t1}) + K_p(Z_{w1} - Z_{t1}) + C_p l_t \dot{\beta}_{t1} + K_p l_t \beta_{t1} = M_w g - 2p_1(t) \\
&M_w\ddot{Z}_{w2} + C_p(\dot{Z}_{w2} - \dot{Z}_{t1}) + K_p(Z_{w2} - Z_{t1}) + C_p l_t \dot{\beta}_{t1} + K_p l_t \beta_{t1} = M_w g - 2p_2(t) \\
&M_w\ddot{Z}_{w3} + C_p(\dot{Z}_{w3} - \dot{Z}_{t2}) + K_p(Z_{w3} - Z_{t2}) + C_p l_t \dot{\beta}_{t2} + K_p l_t \beta_{t2} = M_w g - 2p_3(t) \\
&M_w\ddot{Z}_{w4} + C_p(\dot{Z}_{w4} - \dot{Z}_{t2}) + K_p(Z_{w4} - Z_{t2}) + C_p l_t \dot{\beta}_{t2} + K_p l_t \beta_{t2} = M_w g - 2p_4(t) \\
&\ddot{q}_k(t) + \sum_{i=1}^{N} C_p Z_k(x_i)\sum_{h=1}^{NM} Z_h(x_i)\dot{q}_h(t) + \frac{EI_Y}{m_r}\left(\frac{k\pi}{l}\right)^4 q_k(t) \\
&\quad + \sum_{i=1}^{N} K_p Z_k(x_i)\sum_{h=1}^{NM} Z_h(x_i) q_h(t) - \sum_{i=1}^{N} C_p Z_k(x_i)\dot{Z}_{si}(t) + K_p Z_k(x_i) Z_{si}(t) \\
&= \sum_{j=1}^{4} p_j(t) Z_k(x_{wj}) \quad (k = 1 \sim NM)
\end{aligned} \quad (28)$$

where the wheel-rail force can be computed as

$$p_i(t) = \begin{cases} \left\{\frac{1}{G}\left[Z_{wj}(t) - Z_r(x_{wj},t) - Irre_i(t)\right]\right\} \\ 0 \quad (\text{When the wheel and rail is disengaged}) \end{cases} \quad (29)$$

$$Z_r(x,t) = \sum_{k=1}^{NM} Z_k(x) q_k(t), \quad Z_k(x) = \sqrt{\frac{2}{m_r l}} \sin\frac{k\pi x}{l}$$

NM in Eq. (28) usually needs to be in the hundreds to ensure accuracy, making the datasets too large to handle. Therefore, we let

PINO-MBD: Physics-informed Neural Operator for Solving
Coupled ODEs in Multi-body Dynamics

the neural operator output only the responses of rail under wheelsets:

$$Z_r(x_{wj}, t) = \sum_{k=1}^{NM} Z_k(x_{wj}) q_k(t), \quad j = 1 \sim 4 \quad (30)$$

In this way, the output DOFs of the flexible beam are reduced from NM to 4 and the total DOFs of the system to 14. The drawback of this action is that the full mode information of the rail is not preserved, so the response at an arbitrary position on the rail cannot be reconstructed. Fortunately, the focus of the VTCD is on the responses of the vehicle system as they reflect the running safety and the comfort of passengers. Also, rail vibrations under the wheelsets tend to be the worst, making sure this approach is conservative.

Apart from gravity, the irregularities *Irre* on the rail surface are the main excitations for the VTCD. On real lines, the irregularities are highly random and can be affected by many factors. Random irregularities force the system to vibrate randomly, affecting running safety and riding comfort. The power spectrum density (PSD) function is the most commonly used approach to describe rail irregularity as a stationary random process. In engineering, PSD function is often used to describe the spectrum density concerning frequency. The track irregularity power spectrum is a continuous curve with a spectral density as the ordinate and frequency or wavelength as the abscissa, which can only be obtained through a large number of actual measurements. With the PSD curve, one can generate simulated rail irregularity samples through many inversion methods. In this paper, we use the classic spectrum method [23] to generate irregularity samples (Figure 7 (b)). This method ensures the PSD of the generated samples is identical to the field PSD. Chinese high-speed ballastless track PSD is chosen as the field PSD [23], each sample in the data pair is different.

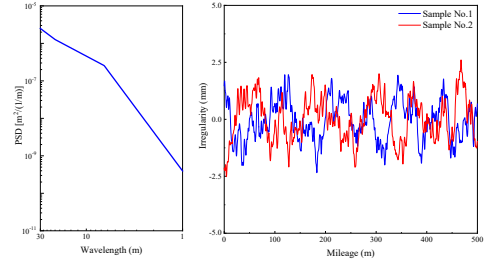

(a). PSD curve  (b). simulated irregularity samples

Figure 7: Chinese high-speed ballastless track PSD and simulated samples with spectrum method

To sum up, we train a PINO-MBD whose input is the VTCD parameters and the output is the system response. The blue symbols in Table 2 are the variable parameters (13 in total). The variation range of each parameter is set to 80%~120% of standard CRH380 high-speed train parameter except for the irregularity. We randomly generate 20000 pairs of data as the training set through Zhai-method [24], and another 2000 pairs as the validation set.

### 4.2 Operator learning with ODEs losses and direct derivatives losses

We train the PINO-MBD on a GTX 1660 card with 6GB of memory. The channel number for each Fourier convolution layer and fully connected layer is set as 72. The initial learning rate is set as 5e-4 and is reduced to 75% every 30 steps. We compute the relative L2 error of solutions, first derivatives, and second derivatives on the validation set at each epoch.

Table 3: Analysis description

| Algorithm | Description | Depth |
|---|---|---|
| 1 | FNO (without physics embedding) | 3 |
| 2 | PINO (without ODE magnitude weight factors) | 3 |
| 3 | PINO-MBD with ODEs losses (with ODE magnitude weight factors) | 3 |
| 4 | PINO-MBD with direct derivatives losses | 3 |
| 5 | PINO-MBD with direct derivatives losses | 5 |

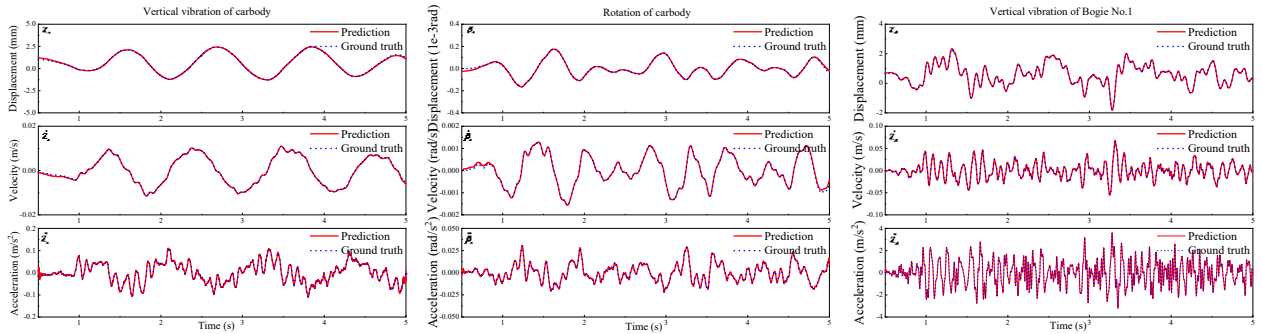

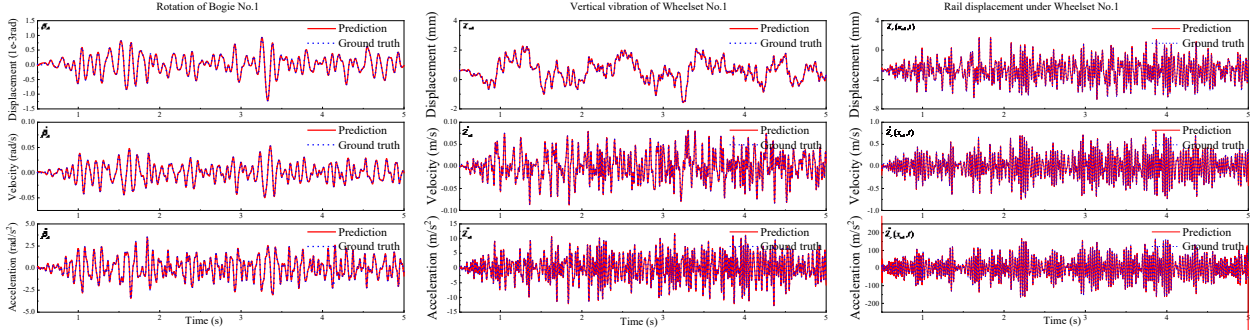

**Figure 8: Performance of PINO-MBD for Train-track coupled dynamics (Algorithm 5)**

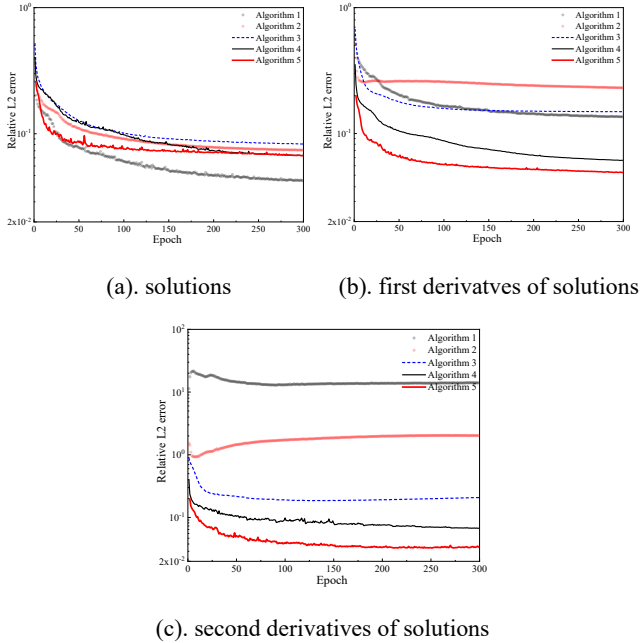

(a). solutions  (b). first derivatves of solutions

(c). second derivatives of solutions

**Figure 9: Influence of physical equation embedding methods on model performance**

The necessity of using embedded physical system equations for the MBD problem can be seen in Figure 9. FNO achieves the best performance on the solutions but performed poorly on first and second derivatives. As mentioned earlier, solution derivatives are usually more important for MBD problems.

**Table 4: Model performance (relative L2 error (%))**

| Algorithm | solutions | First derivatives | Second derivatives |
|---|---|---|---|
| 1 | 4.24 | 13.58 | 1404.42 |
| 2 | 7.41 | 23.06 | 202.42 |
| 3 | 8.29 | 14.90 | 20.79 |
| 4 | 6.75 | 6.12 | 6.73 |
| 5 | **6.77** | **4.90** | **3.20** |

Table 4 shows the relative L2 losses of solutions, first derivatives, and second derivatives for different algorithms. Algorithm 2 (PINO) only configures one weight for the ODEs loss as a whole in the total loss function. As mentioned in section 3.2, neural operator cannot harvest gradients on ODEs with small ODE weight factors. Therefore, the performance on the first and second derivatives of algorithm 2 is unacceptable even with fine-tune. On the contrary, Algorithm 3 takes ODE weight factors into account, so the neural operator can harvest gradients of similar orders of magnitude on different ODEs, so as to achieve better performance.

Algorithm 4 uses the direct gradient loss, and it can be seen that the effect is further improved compared to Algorithm 2. As mentioned in section 3.2.2, ground truth for solution derivatives is usually wasted in existing research. Since it is not difficult to obtain the ground truth of solution derivatives for MBD problems, we recommend using direct gradient loss when one has them. Of course, the downside of this approach is that the file size of training data triples. It is also observed from Figure 10 that the GNN depth has a significant impact on the performance.

In future research, we will utilize better GPUs to optimize PINO-MBD. But in terms of engineering applications, the performance of Algorithm 5 is completely sufficient. Compared with exiting data-driven models in the MBD research field [26-27], PINO-MBD has achieved a significant improvement in accuracy. At the same time, PINO-MBD can make predictions with only a forward pass of the network without additional training, which is not available in most data-driven MBD studies. Compared with the numerical integration algorithm, the improvement in computational efficiency of PINO-MBD is also significant. The simulation ratio for Zhai-method-based solver is approximately 1:25, meaning 1 second of solution requires 25 seconds of simulation. However, the simulation ratio for PINO-MBD is less than 1:0.1. Also, neural operators hardly consume any CPU-like numerical integration methods (over 80%). Generating the training dataset with 20000 data pairs took more than 40 hours on a computer with intel i9-10900X CPU while the trained PINO-MBD needs less than 2 seconds.

## CONCLUSIONS

In this work, we propose the physics-informed neural operator for solving coupled ODEs in MBD (PINO-MBD). With two novel physics-embeddings methods, PINO-MBD achieves satisfying performance on solution derivatives while PINO fails to output acceptable results. The potential engineering and commercial value of PINO-MBD is huge. In a design project, engineers can quickly meet different design requirements by simulating a large number of designs through pre-trained PINO-MBD with almost no computational cost. Furthermore, we can expect PINO-MBD training software with GUI interface to appear in the future, just like the commonly used MBD commercial software Simpack [30], UM [31], ADAMS [32], etc.

## ACKNOWLEDGMENTS

This study was funded by the National Natural Science Foundation of China (NSFC) under Grant Nos. U1934214 and 51878576.